\documentclass[twocolumn, aps, prd]{revtex4}
\usepackage{graphicx}
\usepackage{amssymb}

\begin{document}

\title{Quantum Hall Effect on the Hofstadter Butterfly}
\author{Mikito Koshino and Tsuneya Ando}
\affiliation{
Department of Physics, Tokyo Institute of Technology
2-12-1 Ookayama, Meguro-ku, Tokyo 152-8551, Japan}
\date{\today}

\begin{abstract}
Motivated by recent experimental attempts to detect the Hofstadter
butterfly, we numerically calculate the Hall conductivity in a modulated
two-dimensional electron system with disorder in the quantum Hall
regime.  We identify the critical energies where the states are extended
for each of butterfly subbands, and obtain the trajectory
as a function of the disorder.  
Remarkably, we find that, when the modulation becomes anisotropic,
the critical energy branches accompanying 
a change of the Hall conductivity.
\end{abstract}

\maketitle


The problem of a Bloch electron in a magnetic field in two dimensions has long been investigated.
Theoretically, the interplay of the modulation and the magnetic field
gives rise to a recursive set of energy gaps as a sensitive function of the field amplitude, which is known to be the Hofstadter butterfly \cite{Hofs}.
It is also predicted that the system exhibits the quantum Hall effect when the Fermi energy is in each single gap, where the gap-rich structure leads to the nontrivial sequence of the quantized Hall conductivities \cite{TKNN}.
On the experimental side, many efforts have been devoted to the challenge of observing the Hofstadter butterfly mainly in lateral superlattices patterned on GaAs/AlGaAs heterostructures \cite{Schl,Albr,Geis}.
A remnant of the nonmonotonic behavior of the Hall conductivity peculiar to the Hofstadter butterfly was observed \cite{Albr}.

In a two-dimensional (2D) electron system without periodic potentials,
it is generally believed that the weak disorder makes almost all the
states localized leaving the extended states at the center of the Landau
band, and the Hall conductivity jumps from one integer to another when
the Fermi energy sweeps through the extended states.  It is natural to
expect in a periodic system that each butterfly subband exhibiting a
non-zero Hall conductivity has extended states at specific energies.  It
is an intriguing problem how these energies move and how the
nonmonotonic behavior of the Hall conductivity changes as a function of
disorder strength, until the butterfly structure is eventually
destroyed.

The problems of the quantum Hall effect and Anderson localization 
in the disordered 2D Bloch system have been studied by several authors.
The evolution of the extended states has been investigated
in tight-binding lattices, from the viewpoint of
the disappearance of the quantum Hall effect 
in the weak-field limit \cite{Ando_1989b,Liu,Yang96,Shen97,Hats,Shen01}.
The pair annihilation of extended states with the increase in randomness
was demonstrated \cite{Ando_1989b} and
the Hall conductivity was also calculated \cite{Shen97}.
The nonmonotonic behavior of the Hall conductivity
was demonstrated for the Hofstadter butterfly
in the presence of disorder \cite{Aoki}.  
A finite-size
scaling analysis was performed for a 2D system modulated by a weak
periodic potential and the critical exponent was estimated at the center
of the Landau level \cite{Huck}.  
A qualitative discussion on the evolution of the extended states in
the Hofstadter butterfly as a function of the
disorder for several flux states was
given \cite{Tan,Yang99}.

In this paper, we study the scaling property 
of the Hall conductivity $\sigma_{xy}$ in
a weakly modulated 2D electron system, and identify the critical 
energies for each of butterfly subbands. We obtain the
trajectory in the energy-disorder space to show how the energies
move as a function of disorder.
We also find that the number of critical energies 
sensitively depends on the {\it anisotropy} of the periodic potential,
which is controllable in a real experiment.


Let us consider a two-dimensional system in a strong magnetic field
with a square periodic potential
\begin{equation}
 V_p(x,y) = V_x\cos\frac{2\pi}{a}x + V_y\cos\frac{2\pi}{a}y.
\end{equation}
The band structure is characterized by the parameter $\phi=Ba^2/(h/e)$, 
a number of magnetic flux quanta penetrating a unit cell \cite{Hofs}.
The disorder potential $V_d$ is taken as the
randomly distributed delta potential $\pm v_0$
with the concentration $2\pi l^2 n_i = 4$, where
$l$ is the magnetic length and $n_i$ is the number of
scatterers in a unit area \cite{Ando83}.
The energy scale for the disorder 
is then given by $\Gamma = 4n_i v_0^2/(2\pi l^2)$ \cite{Ando74}.
The magnetic field is assumed to be sufficiently strong
so that $V_p$ and $\Gamma$ are smaller than the cyclotron energy
$\hbar\omega_c$, and so that we can consider only the lowest Landau level.
To deal with a finite system,
we consider the larger unit cell $L\times L$ with $L = Ma$ ($M$: integer)
and apply the periodic boundary condition.
We calculate the Hall conductivity using the Kubo formula
for zero temperature,
\begin{equation}
 \sigma_{xy} = \frac{\hbar e^2}{iL^2} \!\!\!
\sum_{\epsilon_{\alpha} < E_F}\sum_{\epsilon_{\beta} \neq \epsilon_{\alpha}}
\!\!\! \frac{\langle \alpha | v_x | \beta \rangle 
\langle \beta | v_y | \alpha \rangle 
\!-\! \langle \alpha | v_y | \beta \rangle \langle \beta | v_x | \alpha \rangle}
{(\epsilon_{\alpha} - \epsilon_{\beta})^2},
\label{Kubo}
\end{equation}
where $\epsilon_{\alpha}$ is the energy of the eigenstate $|\alpha\rangle$,
$E_F$ the Fermi energy, and $v_i$ the velocity operator.
We take into account the mixing of the states 
between Landau levels $n=0$ and 1
up to the first order of $V/(\hbar\omega_c)$ with $V$ being $V_p$ or
$V_d$. 

\begin{figure}
\begin{center}
\leavevmode\includegraphics[width=70mm]{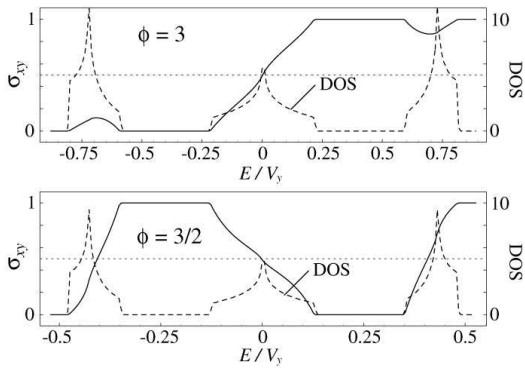}
\end{center}
\caption{Plots of the Hall conductivity $\sigma_{xy}$
(in units of $-e^2/h$) for a nondisordered electron in the modulation
$V_x/V_y=1$ with $\phi = 3$ (upper) and 3/2 (lower).
Dashed lines represent the density of states in units of $1/(V_y a^2)$.
}
\label{fig_noimp}
\end{figure}

\begin{figure}
\begin{center}
\leavevmode\includegraphics[width=70mm]{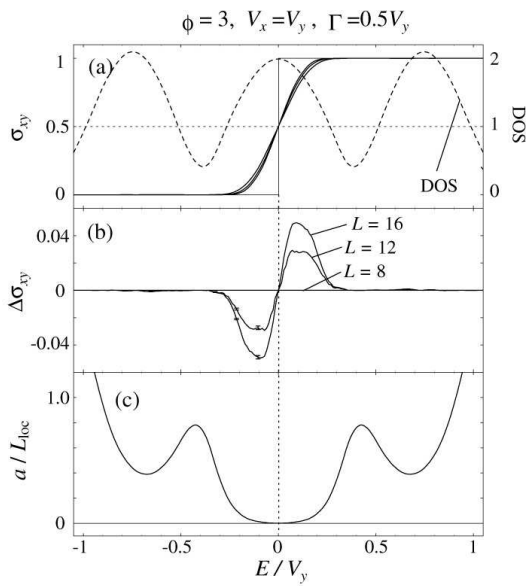}
\end{center}
\caption{
(a) Hall conductivity $\sigma_{xy}$ (in units of $-e^2/h$)
calculated for disordered systems with
$V_x/V_y=1$, $\Gamma/V_y=0.5$, $\phi = 3$,
and $L/a=8$, 12, and 16.
The dashed line represents the density of states in units of $1/(V_y a^2)$.
(b) Relative values of $\sigma_{xy}$ measured from
the smallest ($L/a=8$) sample.
(c) Inverse localization length (in units of $a$)
estimated from the Thouless number. 
The vertical dashed line penetrating the panels 
represents the critical energy.
}
\label{fig_iso3}
\end{figure}

\begin{figure}
\begin{center}
\leavevmode\includegraphics[width=70mm]{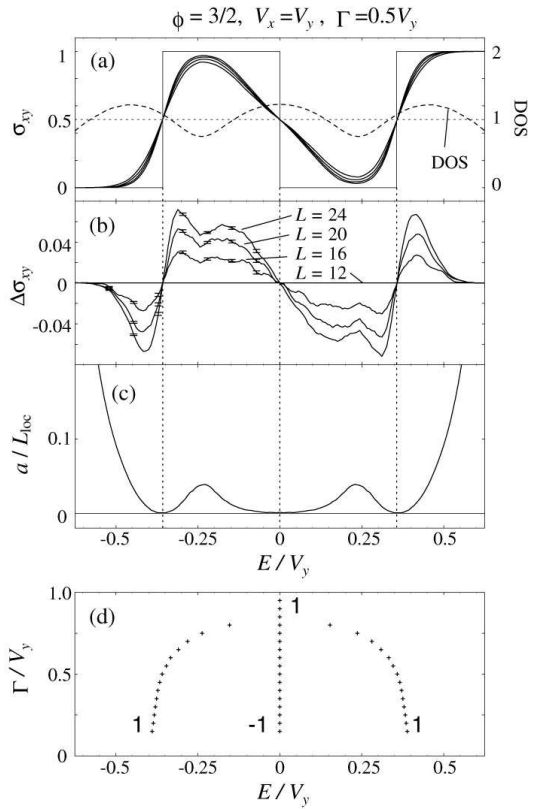}
\end{center}
\caption{Plots similar to Fig.\ \ref{fig_iso3},
for $V_x/V_y=1$, $\Gamma/V_y=0.5$, and $\phi = 3/2$ with $L/a=12$, 16, 20, and 24.
The panel (d) shows the trajectories of the
critical energies as a function of $\Gamma$.
Numbers represent the corresponding Hall conductivities.
}
\label{fig_iso32}
\end{figure}


We first consider an isotropic modulation, $V_x=V_y$.
The calculations are performed for two different fluxes,
$\phi = 3$ and 3/2.
In both cases, the lowest Landau level
splits into three subbands in the absence of the disorder, 
while the distributions
of the Hall conductivity $\sigma_{xy}$ are different 
as shown in Fig.\ \ref{fig_noimp}.
Specifically, $\sigma_{xy}$ for three subbands 
becomes $(0,1,0)$ (in units of $-e^2/h$) in $\phi = 3$ 
and $(1,-1,1)$ in $\phi = 3/2$, which satisfies the
Diophantine equation \cite{TKNN}.

Figures \ref{fig_iso3} and \ref{fig_iso32} show 
the numerical results calculated 
for the disordered systems with several sizes
for $\phi = 3$ and 3/2, respectively.
The panel (a) shows $\sigma_{xy}$ with the density of states,
and (b) the difference in $\sigma_{xy}$
measured from the smallest sample.
We also show in the panel (c)
the inverse localization length $1/L_{\rm loc}$ 
estimated by the Thouless number method \cite{Ando83}.
We divide the energy axis into 200 columns and
every quantity is averaged over a number of different samples
for each column.

We can clearly see that
the subbands with zero Hall conductivity (the upper and lower
subbands in $\phi = 3$)
are all localized as naturally expected,
and the Hall conductivity there
rapidly reaches the quantized value.
Each of other bands contains
a critical energy where the localization length 
diverges ($1/L_{\rm loc} = 0$).
The Hall conductivity for these delocalized bands
has fixed points at $\sigma_{xy} = 1/2$,
which agree quite well with the critical energies estimated 
from $L_{\rm loc}$.
$\sigma_{xy}$ off the fixed points always approaches 1 in the area $\sigma_{xy} > 1/2$ and 0 in the area $\sigma_{xy} < 1/2$, leading to
the Hall plateau in an infinite system (shown as
step like lines in (a)).
Therefore, it is natural to define the positions of the extended
states as the points of $\sigma_{xy}=1/2$, which is exactly
the method adopted in the following.
The panel (d) in Fig.\ \ref{fig_iso32} shows the trace of 
the critical energies 
identified as the positions of $\sigma_{xy}=1/2$
for several values of $\Gamma$.
Three lines with the Hall conductivities 1,-1, and 1 become closer 
as the magnitude of disorder increases, 
and combine into one critical energy with the Hall conductivity 1.


\begin{figure}
\begin{center}
\leavevmode\includegraphics[width=70mm]{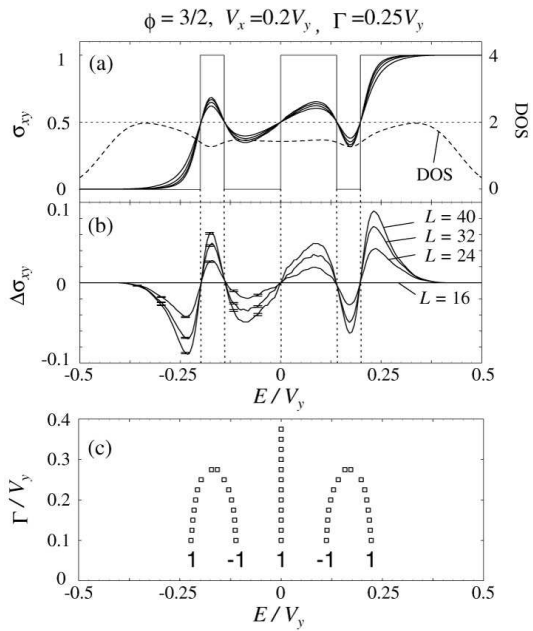}
\end{center}
\caption{
(a) Hall conductivity $\sigma_{xy}$ (in units of $-e^2/h$)
calculated for disordered anisotropic systems with 
$V_x/V_y =0.2$, $\Gamma/V_y = 0.25$, $\phi = 3/2$ 
and $L/a=16$, 24, 32, 40.
The dashed line represents the density of states in units of $1/(V_y a^2)$.
(b) Relative values of $\sigma_{xy}$ measured from
the smallest ($L/a=16$) sample.
(c) Trajectories of the
critical energies as a function of $\Gamma$.
Numbers represent the corresponding Hall conductivities
}
\label{fig_aniso}
\end{figure}

\begin{figure}
\begin{center}
\leavevmode\includegraphics[width=70mm]{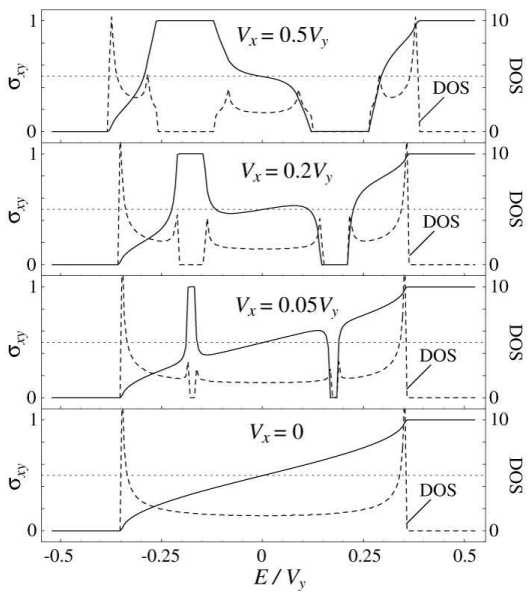}
\end{center}
\caption{Plots of the Hall conductivity $\sigma_{xy}$
(in units of $-e^2/h$)
for a nondisordered electron in anisotropic systems
with $V_x/V_y = 0.5$, 0.2, 0.05 and 0 and $\phi = 3/2$.
The dashed line represents 
the density of states in units of $1/(V_y a^2)$.
}
\label{fig_noimp_ani}
\end{figure}

\begin{figure}
\begin{center}
\leavevmode\includegraphics[width=70mm]{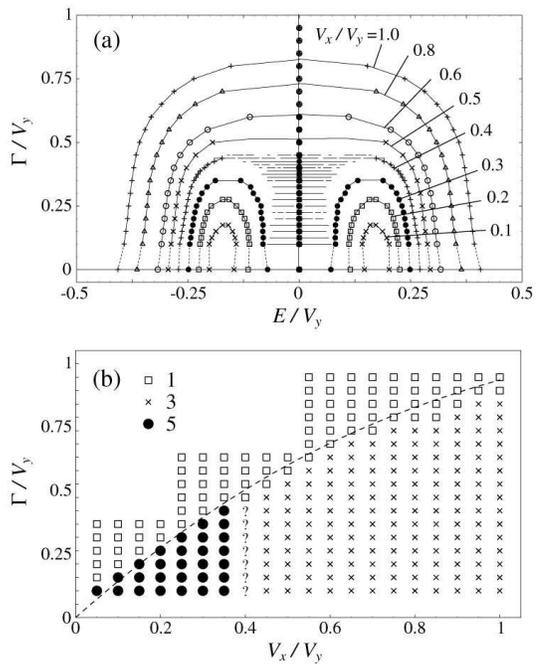}
\end{center}
\caption{(a) Traces of the critical energies 
for several values of $V_x/V_y$.
For $V_x/V_y = 0.4$, we present the
energy regions where the error bar of the Hall conductivity
reaches $\sigma_{xy} = 1/2$.
(b) Phase diagram showing the 
number of critical energies in 
the disorder ($\Gamma/V_y$)-anisotropy ($V_x/V_y$) space.
`?' indicates indeterminate.
The dashed line shows $4E_G/V_y$
with $E_G$ being the subband gap in the ideal system without disorder.
}
\label{fig_phase}
\end{figure}

The localization property is strongly affected by the
configuration of the periodic potential itself.
Here, we investigate how the above results 
are altered when we make the potential {\it anisotropic}
(i.e., $V_x\neq V_y$).
Figure \ref{fig_aniso} shows the Hall conductivity
calculated for the disordered system with $V_x/V_y =0.2$
and $\phi=3/2$.
Remarkably, we found {\it five} fixed points at $\sigma_{xy} = 1/2$,
and they likely lead to the critical energies with
the Hall conductivities 1, $-1$, 1, $-1$, and 1 
in an infinite system, shown as a zig-zag line in (a).
We also calculated the Thouless number,
but failed to estimate the localization length
around the extended region,
since we cannot resolve the exponent 
of the size dependence of the Thouless number.
However, the robust fixed points and the
monotonic behavior with $L$ in the Hall conductivity
strongly suggests the existence of the critical energies
at those points.
The trajectories of the fixed points
for several values of $\Gamma$ are shown in Fig.\ \ref{fig_aniso}(c).
At a certain point, the upper and lower pairs annihilate
and the center is left,
which is topologically different from the isotropic case.

The origin of the five critical energies can be
explained as follows.
Figure \ref{fig_noimp_ani} 
shows the Hall conductivity $\sigma_{xy}(E)$ for clean systems
with the fixed flux $\phi=3/2$ and various ratios of $V_x/V_y$.
The gaps between subbands rapidly shrink
as $V_x/V_y$ goes apart from 1 
and disappear at $V_x=0$ (1D modulation), 
where $\sigma_{xy}(E)$ behaves just
monotonically from 0 to 1.
When a very small $V_x$ is present, we see that 
$\sigma_{xy}(E)$ is deviated from that in $V_x=0$
only around the subband gaps ($E-E_{\rm gap} \lesssim V_x$).
There the perturbative $V_x$ mixes
the states in $V_x=0$ to produce positive and negative Hall currents
so that the Hall conductivity is quantized in the gap.
If a small but finite disorder is introduced into this system, 
the critical energies with opposite Hall conductivities
should exist correspondingly on each side of the gap,
as long as $\Gamma$ is smaller than the gap width
and the Hall plateau survives.
The electron-hole symmetry
requires another critical energy right at the Landau band center.
When $V_x$ is switched off, the critical energy pairs at the gaps 
annihilate and the central one is left.

We show the corresponding plots for various values 
of $V_x/V_y$ in Fig.\ \ref{fig_phase}(a), 
to explain how the isotropic (three critical energies)
and anisotropic cases (five energies) are connected. 
We see that the center branch in the isotropic case
splits into three branches at $V_x/V_y \approx 0.4$.
The points at $\Gamma = 0$ show the energies 
of $\sigma_{xy} = 1/2$ in clean systems.
The critical energies should converge 
to these points in the limit $\Gamma \rightarrow 0$,
as long as $\sigma_{xy} = 1/2$ is the critical point for
the disordered systems with a finite size.
The number of such points changes from 3 to 5
at $V_x/V_y=(V_x/V_y)_0 \approx 0.36$, which defines the turning point
in the evolution of the critical energy.
The results (although not included in Fig.\ \ref{fig_phase}(a)) show
that, for $V_x/V_y=0.35<(V_x/V_y)_0$, there are five critical energies
when $\Gamma$ is sufficiently small, and for $V_x/V_y=0.45$, there are
only three critical energies.  In the case $V_x/V_y = 0.4$, however, it
is difficult to resolve the critical energies around $E=0$, partly
because of the numerical accuracy of the calculated $\sigma_{xy}$.
Therefore, we can present only the energy regions where the error bar of
the Hall conductivity reaches $\sigma_{xy} = 1/2$.

We give the phase diagram in Fig.\ \ref{fig_phase}(b), showing the
number of critical energies in the disorder ($\Gamma/V_y$)-anisotropy
($V_x/V_y$) space.  The newly found five-energy phase occupies a
considerable area in the region of high anisotropy and low disorder.
The dashed line shows $4E_G/V_y$ as a function of $V_x/V_y$, where $E_G$ is
the subband gap in the ideal system without disorder.  The critical
disorder where the number of critical energies changes from five to one
or three to one roughly follows the curve as might be expected.

While we concentrated on the flux $\phi = 3/2$ here,
it is expected that, in other fluxes,
the butterfly gaps similarly shrink as the modulation goes 
to 1D and the pair annihilation of
the extended states occurs at the butterfly gaps.
We also assume that such a behavior 
does not basically change in higher Landau levels.
We expect that controlling the QHE by the anisotropy
proposed here is sufficiently promising for a real experiment,
if a modulation is given by the external gate electrodes
as in the recent experiment \cite{Albr}.
The existence of the fixed point at $\sigma_{xy} = 1/2$ 
is compatible with the the two-parameter scaling theory \cite{Prui},
and investigating the 
$(\sigma_{xy},\sigma_{xx})$ scaling in the periodic system
is an important future work.

%
This work has been supported in part by a 21st Century COE Program at
Tokyo Tech \lq\lq Nanometer-Scale Quantum Physics'', a Grant-in-Aid
for COE (12CE2004 \lq\lq Control of Electrons by Quantum Dot Structures
and Its Application to Advanced Electronics''),
and Scientific Reseach from the Ministry of
Education, Culture, Sports, and Technology, Japan. 
Numerical calculations were
performed in part using the facilities of the Supercomputer Center,
Institute for Solid State Physics, University of Tokyo.  \par
%

\end{document}